\documentclass[prl,aps,twocolumn,showpacs,superscriptaddress,notitlepage,nofootinbib]{revtex4-1}

\usepackage{graphicx}
\usepackage{ae,aecompl}
\usepackage{amsmath}
\usepackage{amssymb}
\usepackage{times}
\usepackage{color}
\usepackage{bbm}
\usepackage{bm}
\usepackage{xcolor}
\usepackage{soul}

\usepackage{textcomp}
\usepackage{soul}
\usepackage{bbold}
\usepackage{pdfpages}

\usepackage{anysize}
\marginsize{1.4cm}{1.4cm}{2.8cm}{2.3cm}

\usepackage[colorinlistoftodos, shadow,textsize= footnotesize]{todonotes}

\newcommand{\new}[1]{\textcolor{black}{#1}}

\newcommand{\mean}[1]{\ensuremath{\langle #1 \rangle}}

\newcommand{\vect}[1]{\bm{#1}}
\newcommand{\be}{\begin{equation}}
\newcommand{\ee}{\end{equation}}

\newcommand{\beq}{\begin{eqnarray}}
\newcommand{\eeq}{\end{eqnarray}}

\newcommand{\ket}[1]{\ensuremath{|#1\rangle}}

\newcommand{\Po}{P_0}
\newcommand{\Pd}{P_{\delta \theta}}
\newcommand{\eps}{\varepsilon}

\newcommand{\vpr}{\upsilon_{\rm Q}^2}
\newcommand{\vQ}{\upsilon_{\rm Q}^2}
\newcommand{\vmax}{\upsilon_{\rm max}^2}

\begin{document}

\title{Witnessing Entanglement without Entanglement Witness Operators}

\author{Luca Pezz\`e}
\affiliation{QSTAR, INO-CNR and LENS, Largo Enrico Fermi 2, 50125 Firenze, Italy}

\author{Yan Li}
\affiliation{Institute of Theoretical Physics and Department of
Physics, Shanxi University, 030006, Taiyuan, China}

\author{Weidong Li}
\affiliation{Institute of Theoretical Physics and Department of
Physics, Shanxi University, 030006, Taiyuan, China}

\author{Augusto Smerzi}
\affiliation{QSTAR, INO-CNR and LENS, Largo Enrico Fermi 2, 50125 Firenze, Italy}

\begin{abstract} 
Quantum mechanics predicts the existence of correlations between composite systems 
that, while puzzling our physical intuition, 
enable technologies not accessible in a classical world. 
Notwithstanding, there is still no efficient general method 
to theoretically quantify and experimentally detect entanglement of many qubits.
Here we propose to detect entanglement by measuring the statistical response of a 
quantum systems to an arbitrary nonlinear parametric evolution. 
As a major difference with respect to current approaches based on the implementation of
entanglement witness operators,
we witness entanglement without relying on measurement efficiencies
or tomographic reconstructions of the quantum state.
The protocol requires only two collective settings for any number of parties.
To illustrate its user-friendliness 
we demonstrate multipartite entanglement in different experiments with ions and photons by analyzing 
published data on fidelity visibilities and variances of collective observables.
\end{abstract}

\date{\today}

\maketitle

A central problem in quantum technologies is to detect and characterize entanglement among correlated parties
\cite{GuhnePHYSREP2009, HorodeckiRMP2009, AmicoRMP2008}. 
On the experimental side, the challenge is to certifty entanglement in presence of
imperfect measurements and decoherence due to the interaction of the system with the environment. 
A most popular approach is based on the implementation 
of entanglement witness operators (EWOs). 
An EWO is a Hermitian operator $\mathcal{W}$ 
such that $\mathrm{Tr}[\rho_{\rm sep} \mathcal{W}] \geq 0$ for all separable states $\rho_{\rm sep}$, and 
$\mathrm{Tr}[\rho \mathcal{W}]<0$ for, at least, one 
entangled state $\rho$~\cite{HorodeckiJPA1996, HorodeckiPLA2001, TerhalTHCSCI2002, LewensteinPRA2000, VogelPRL2013}.
The power of this method relies on the algebraic fact that for 
each multipartite entangled state it exists (at least) one EWO that recognizes it \cite{HorodeckiJPA1996}. 
The experimental protocol implements projective measurements with the eigenvectors of $\mathcal{W}$ so to
directly extract $\mathrm{Tr}[\rho \mathcal{W}]$.
A drawback is that EWOs are device-dependent: they require precise and non-trivial assumptions
on the efficiencies and fidelities of the projective measurements. 
In practice, experimental imperfections may easily lead to false positives, namely, to 
the unwitting realization of operators $\mathcal{W}_{\rm exp} \neq \mathcal{W}$
such that $\mathrm{Tr}[\rho_{\rm sep} \mathcal{W}_{\rm exp}]<0$ also for (some) separable states, therefore 
signaling entanglement in states that are only classically correlated \cite{SeevinckPRA2001, RossetPRA2012}. 
The same problem arises, even more dramatically, when trying to detect entanglement via the 
tomographic reconstruction of the quantum state \cite{LvovskyRMP2009, RossetPRA2012, MoroderPRL2013}.
This problem has spurred an intense search for more robust entanglement witnesses criteria \cite{Adesso, LoredoPRL2016}. 
 
A reliable experimental detection of entanglement requires  
the implementation of device-independent witness operators. 
An important class of entangled states is recognized by Bell-like inequalities
testing the correlations between measurement data (obtained for different settings of non-communicating parties).
These correlations among noninteracting parties cannot be created by local operations 
and experimental imperfections. 
Bell-like tests can thus detect entanglement without 
relying on any hypothesis about the measurement actually performed,
nor they need any specific assumption on the state like, for instance, its Hilbert space dimension.
Therefore, it has been suggested that Bell-like inequalities are device-independent entanglement 
witness (DIEW) operators~\cite{BancalPRL2011}
and have been applied to maximally entangled
states~\cite{NagataPRL2002, SeevinckPRL2002, BancalPRL2011, PalPRA2011} of $N$ qubit systems.
Generally speaking, DIEWs demand that the parties i) must be addressed locally and 
ii) do not interact during the local operations and measurements.
Also, the choice of the local operations settings 
is not straightforward: the specific configurations required to witness entanglement are 
only known for particular cases.  The extension to an arbitrary state can vary from
computationally hard to prohibitive since it can increase exponentially with the number of qubits. 
A recent experiment with trapped ions~\cite{BarreiroNATPHYS2013} has exploited Bell-based DIEWs 
and demonstrated genuine multipartite entanglement up to 6 particles.
Crosstalk among the parties affected the DIEW of larger systems~\cite{BarreiroNATPHYS2013}. 

\begin{figure}[t!]
\includegraphics[width=\columnwidth]{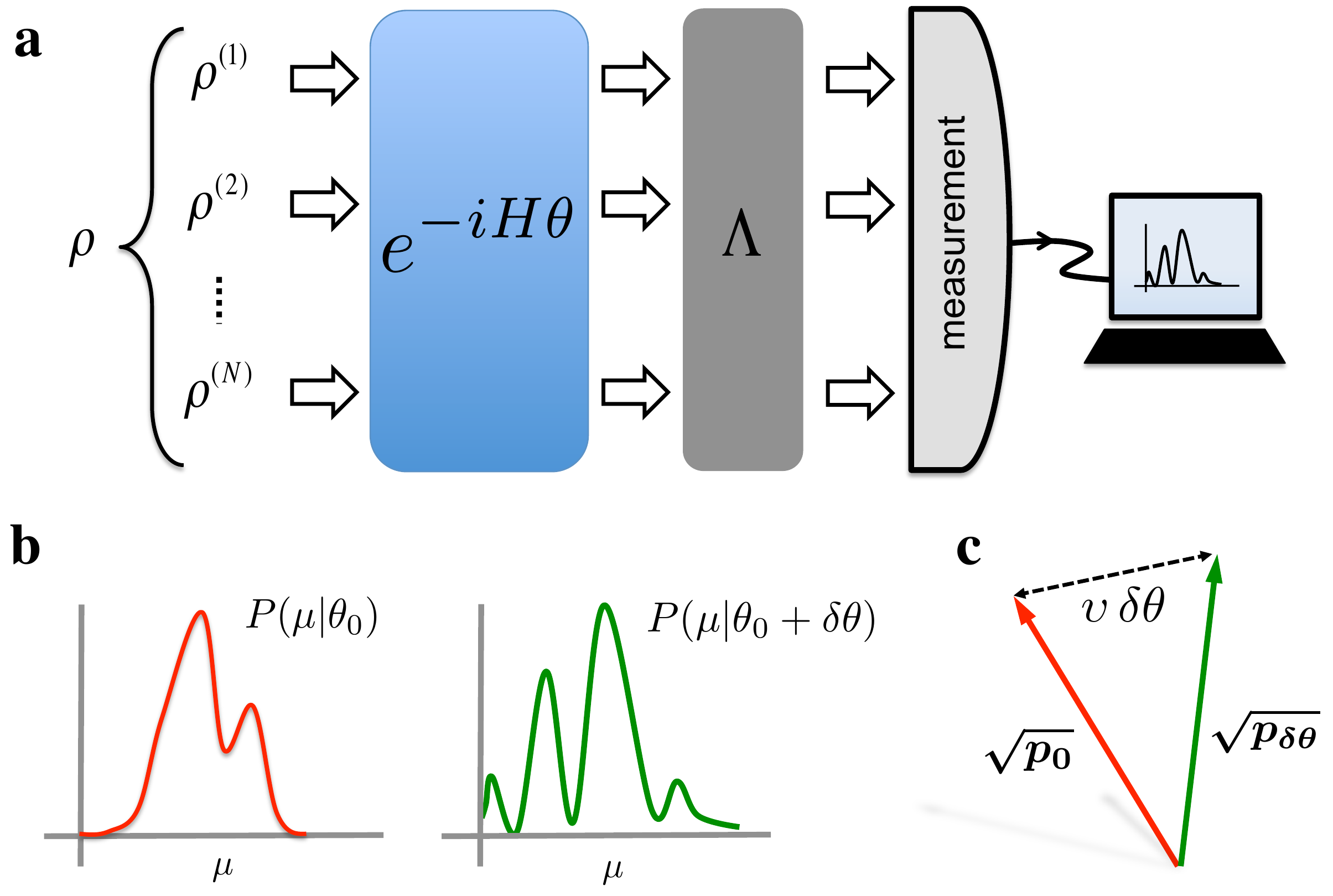}
\caption{\textbf{Statistical speed.} 
({\bf a}) $N$ parties prepared in a quantum state $\rho$ undergo
a unitary transformation with $\theta$ a tunable parameter.  
The map $\Lambda$ includes arbitrary $\theta$-independent decoherence effects. 
({\bf b}) The probability distribution $P(\mu \vert \theta)$
is obtained by collecting the measurement results $\mu$ for different values of the parameter that here  
are chosen to be $\theta=\theta_0$ (red line) and $\theta=\theta_0+\delta \theta$ (green line).
({\bf c}). To quantify the statistical distinguishability between the two distributions
we introduce unit vectors $\bm{\sqrt{p_0}} = \{ \protect\sqrt{P(\protect\mu \vert \theta_0)} \}_\protect\mu$ (red) and 
$\bm{\sqrt{p_{\theta}}} = \{ \protect\sqrt{P(\protect\mu\vert \protect\theta)} \}_\protect\mu$ (green) and 
measure the Euclidean distance among them: $\new{\ell} \equiv 2 \vert\vert \bm{\sqrt{p_0}} - \bm{\sqrt{p_{\theta}}} \vert\vert$ (dashed line). 
The statistical speed $\upsilon = \tfrac{d \new{\ell}}{d \theta}\vert_{\theta_0}$ is an entanglement witness. }
\label{fig1}
\end{figure}

In this manuscript we propose a novel approach to witness entanglement
based on the statistical speed of a quantum system driven by an arbitrary nonlinear transformations, see Fig.~1.
As in the case of Bell-based DIEWs, it is state-independent and free from any constraint 
on the measurement efficiencies.
However, our method does not require local manipulations 
and measurements: it detects entanglement with collective transformations even in presence of cross-talking between any number of qubits. 
It also extends the detection of entanglement beyond the usual realm of Bell-like tests to general multivariate observables.  
In the particular case of linear transformations, the statistical speed detects 
$k$-partite entanglement~\cite{PezzePRL2009, TothPRA2012, HyllusPRA2012}. 
As it will be explained in detail below, our approach is not fully device independent because it requires the
experimental control of the collective transformation. 
Several experiments have shown the feasibility of controlled collective phase shifts
with cold \cite{AppelPNAS2009,BohnetNATPHOT2014} and ultracold atoms \cite{StrobelSCIENCE2014, LuckeSCIENCE2011, RiedelNATURE2010}, 
ions \cite{HaffnerNATURE2005, MonzPRL2011,LeibfriedSCIENCE2005,LeibfriedNATURE2006, Bohnet_2015}, photons \cite{GaoNATPHYS2012,YaoNATPHOT2012} 
and superconducting circuits \cite{DiCarloNATURE2010}.
We therefore demonstrate, by just elaborating on published 
experimental data \cite{WaltherNATURE2004, LeibfriedSCIENCE2005, LeibfriedNATURE2006, GaoNATPHYS2012}, 
entanglement up to 14 ions and 10 photons, and genuine multipartite entanglement up to 6 ions -- in agreement with Bell-based DIEW results reported in Refs.~\cite{BarreiroNATPHYS2013}. 
This hallmarks the simplicity and interdisciplinary nature of our approach.


\section{Results}

{\bf Witnessing entanglement via a statistical speed.}
Figure~1 illustrates the basic ingredients of our entanglement witness protocol.
A quantum state $\rho$ is probed by applying a collective transformation parametrized by a real number $\theta$.
The output state is characterized by the statistical probability distribution $P(\mu \vert \theta)$ of possible measurement results $\mu$ for a generic observable.
The distinguishability between the two probability distributions $P(\mu \vert \theta_0)$ and
\new{$P(\mu \vert \theta)$} is quantified by the 
Hellinger distance \cite{BZBook}
\begin{equation}  \label{Eq.Hellinger}
\new{\ell(\theta_0, \theta) = 2 \sqrt{ \sum_\mu \Big( \sqrt{P(\mu \vert \theta_0)} - \sqrt{P(\mu\vert \theta)} \Big)^2} },
\end{equation}
with the sum extending over all possible measurement results $\mu$.
\new{$\ell$} is a statistical distance \cite{WoottersPRD1981, BraunsteinPRL1994}: it ranges from zero, if and only if 
\new{$P(\mu \vert \theta_0) = P(\mu\vert \theta)$}
$\forall \mu$, to its maximum value $\new{\ell} = 2\sqrt{2}$, 
if and only if 
\new{$P(\mu \vert \theta_0) \times P(\mu\vert \theta) =0$}
$\forall \mu$, 
and it satisfies the triangular inequality. 
The Hellinger distance is proportional to the Euclidean
distance 
\new{$\vert\vert \bm{\sqrt{p_0}} - \bm{\sqrt{p_{\theta}}} \vert\vert$}
between the unit vectors $\bm{\sqrt{p_0}} = \{ \sqrt{P(\mu\vert \theta_0)} \}_\mu$ and 
\new{$\bm{\sqrt{p_{\theta}}} = \{ \sqrt{P(\mu\vert \theta)} \}_\mu$}. 
It is useful to introduce the notion of statistical speed: 
$\upsilon \equiv \upsilon(\theta_0) = \tfrac{d \new{\ell}}{d \theta} \vert_{\theta_0}$, i.e. the rate at which \new{$\ell(\theta_0, \theta)$} 
changes with $\theta$ around the reference point $\theta_0$.
A Taylor expansion of Eq.~(\ref{Eq.Hellinger}) gives
\be \label{Eq.Fisher}
\upsilon^2 = \sum_{\mu} \frac{1}{P(\mu \vert \theta_0)} \bigg( \frac{d P(\mu\vert \theta)}{d \theta} \Big\vert_{\theta_0} \bigg)^2.
\ee
The squared statistical speed coincides with the ``classical'' Fisher information~\cite{HelstromBOOK}.
The specific measurement observable entering in Eqs.~(\ref{Eq.Hellinger}) and~(\ref{Eq.Fisher})
via the conditional probabilities is arbitrary but, in practice, chosen so to efficiency distinguish the two probability distributions.   
Extracting the statistical speed requires (at least) two settings, independently from the number of particles, the quantum state and the measurement observable.
Equation~(\ref{Eq.Fisher}) is bounded by the quantum statistical speed 
$\upsilon^2 \leq \vQ =
 {\rm Tr}[\rho L^2]$, where $L$ is the symmetric logarithmic derivative (SLD) uniquely defined on the support of 
$\rho$ via the relation $\tfrac{d \rho}{d \theta} = \tfrac{L \rho + \rho L}{2}$ \cite{BraunsteinPRL1994, HelstromBOOK}.
For any quantum state, the bound can be saturated by optimal measurements \cite{BraunsteinPRL1994}.
Recently, it has been shown that the quantum statistical speed is linked to to the dynamic susceptibility \cite{Hauke_2016} and it is thus readily available in condensed-matter experiments.

Let us consider $N$ qubits and apply the unitary transformation $e^{-i H \theta}$~(see Fig.~1), where   
\be \label{H0H1}
H = \sum_{i=1}^N \frac{\alpha_i}{2} \sigma^{(i)}_{\vect{m}} + \eps \sum_{\substack{i,j=1 }}^N \frac{V_{ij}}{4} \sigma^{(i)}_{\vect{n}} \sigma^{(j)}_{\vect{n}},
\ee
coefficients $\alpha_i$ (without loss of generality, $0 \leq \alpha_i \leq 1$) account for (possibly) inhomogeneous linear couplings
or local/subgroups operations on the parties, $V_{ij}=V_{ji}$ and $\eps$ is an arbitrary real number.
Here, $\sigma^{(i)}_{\vect{n}} \equiv \hat{\vect{\sigma}}^{(i)}\cdot \vect{n}$ 
is the Pauli matrix 
for the $i$th particle and $\vect{n}$ is a versor.
What is the highest statistical speed obtained over all classically correlated states?
The bound 
\be \label{Eq.statspeed}
\upsilon^2(\eps) \leq \max_{\ket{\psi_{\rm pr}}} \vQ (\eps) \equiv \vmax(\eps)
\ee
holds, where the maximum of the quantum statistical speed is taken over all pure product state, $\ket{\psi_{\rm pr}} = \ket{\psi^{(1)}} \otimes ... \otimes \ket{\psi^{(N)}}$.
As derived in Appendix, the quantum statistical speed of $\ket{\psi_{\rm pr}}$ probed by the Hamiltonian $H$ is  
\be \label{Eq.vH}
\vQ(\eps) = \upsilon_{0}^2 + \eps \, \upsilon_1^2 + \eps^2 \,  \upsilon_{2}^2, 
\ee
where 
\beq 
\upsilon_0^2 &=& \sum_{i=1}^N \alpha_i^2 (1 - \mean{\sigma^{(i)}_{\vect{m}}}^2),  \\
\upsilon_1^2 &=& 2 \sum_{\substack{i,j=1 \\ i\neq j}}^N V_{ij} \alpha_i \big[ \vect{n} \cdot \vect{m} - \mean{\sigma^{(i)}_{\vect{n}}} \mean{\sigma^{(i)}_{\vect{m}}} \big] \mean{\sigma^{(j)}_{\vect{n}}},
\eeq
and 
\be \label{Eq.vH1}
\begin{split}
& \upsilon_2^2
= \sum_{\substack{i,j=1 \\ i\neq j}}^N \frac{V_{ij}^2}{2} \big[1- \mean{\sigma^{(i)}_{\vect{n}}}^2\mean{\sigma^{(j)}_{\vect{n}}}^2\big] + \\
&\qquad \qquad \,\,\,\,\, + \sum_{\substack{i,j,l=1 \\ i \neq j \neq l}}^N V_{ij} V_{il} \big[1- \mean{\sigma^{(i)}_{\vect{n}}}^2\big] \mean{\sigma^{(j)}_{\vect{n}}} \mean{\sigma^{(l)}_{\vect{n}}}.
\end{split}
\ee
Because of the convexity of the Fisher information \cite{PezzePRL2009}, 
the bound $\upsilon^2(\eps) \leq  \vmax(\eps)$ holds not only for pure states but also for any statistical mixture of product states 
(i.e. for an arbitrary classically-correlated state).
 
As $\upsilon_{\rm max}(\eps)$ bounds the statistical speed over all possible observables and all separable states, 
states violating the inequality (\ref{Eq.statspeed}) are entangled. 
For linear Hamiltonians ($\eps=0$) the maximization is readily done:
the optimal states have $\mean{\sigma^{(i)}_{\vect{m}}}=0$ $\forall i$, giving $\upsilon_{\rm max}^2(0) = \sum_{i=1}^N \alpha_i^2$.
This generalizes the bound $\upsilon_{\rm max}^2(0)=N$ discussed in \cite{PezzePRL2009} for homogeneous coupling $\alpha_i=1$.
For nonlinear Hamiltonians ($\eps \neq 0$) the bound depends on the explicit form of $V_{ij}$ and $\alpha_i$.
It can be calculated either numerically or, as shown below, analytically in many cases of interest.
In the following we consider, as an example, the Ising model having nearest-neighbor interaction $V_{ij}=\tfrac{\delta_{j,i+1} + \delta_{j,i-1}}{2}$
and $\vect{n}\cdot \vect{m}=1$. 
In Appendix we also report the results for the Lipkin-Meshkov-Glick (LMG) model where $V_{ij}=1$.

\begin{figure}[t!]
\includegraphics[width=\columnwidth]{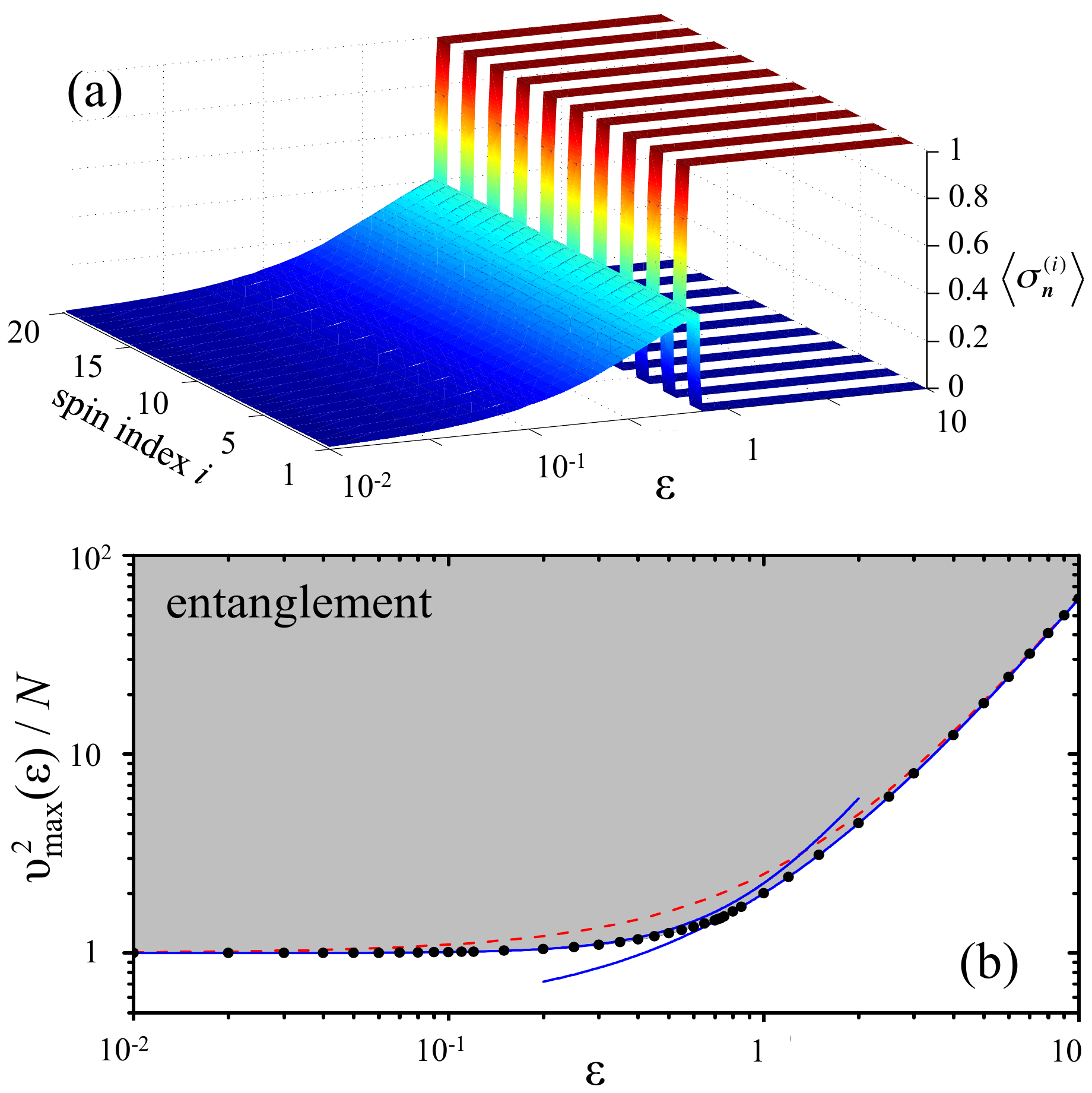}
\caption{\textbf{Witness of entanglement with the Ising Hamiltonian.} 
({\bf a}) Mean spin values $\mean{\sigma_{\vect{n}}^{(i)}}$ for the separable states maximizing $\vpr$, as a function of $\eps$.
({\bf b}) Maximum statistical speed of separable states, $\vmax$ (dots), probed by the Ising Hamiltonian.  
Entanglement is witnessed by a statistical speed $\upsilon^2 > \vmax$, i.e. in the grey region. 
The solid blue lines are analytical limits discussed in the main text. 
The dashed line is the upper bound Eq.~(\ref{Eq.boundNN}).
}
\label{fig2}
\end{figure}

The states that maximize Eq.~(\ref{Eq.vH}) are 
\be
|\psi(\eps) \rangle =\prod_{i=1}^N \sqrt{\frac{1+ \mean{ \sigma_{\vect{n}}^{(i)} } }{2}}~\ket{\uparrow}_i 
+ e^{-i\varphi_i} \sqrt{\frac{1- \mean{ \sigma_{\vect{n}}^{(i)} } }{2}}~\ket{\downarrow}_i
\ee
where $\ket{\uparrow}$ and $\ket{\downarrow}$ are eigenstates of $\sigma_{\vect{n}}$, $\varphi_i$ are arbitrary phases and 
$\mean{\sigma^{(i)}_{\vect{n}}}$ as a function of $\eps$ is reported in Fig.~\ref{fig2}(a).
Figure~\ref{fig2}(b) shows $\vmax(\eps)$ as a function of $\eps$.
The numerical analysis in the homogeneous case $\alpha_i=1$ reveals that, for $\eps$ smaller than a critical value $\eps_c$, 
the speed $\vpr$ is maximized when $\mean{\sigma^{(i)}_{\vect{n}}}$ are all equal. 
In particular, for $\eps \ll 1$, $\vpr$ is the highest when $\mean{\sigma_{\vect{n}}^{(i)}} = \eps$ $\forall i$, giving
$\frac{\vmax(\eps)}{N} = 1+\frac{5}{4}\eps^2 + O(\eps^4)$ (solid blue line in Fig.~\ref{fig2}b).
The value $\eps_c= 0.7302$ is found analytically, as discussed in Appendix. 
For $\eps > \eps_c$,  $\vpr(\eps)$ is maximized by alternating 
$\mean{ \sigma_{\vect{n}}^{(i)} } = 1$ and $\mean{ \sigma_{\vect{n}}^{(i+1)} }= 0$.
In this limit, we find $\frac{\vmax(\varepsilon)}{N} = \frac{1}{2} + \eps +\frac{1}{2}\eps^2$ (solid blue line in Fig.~\ref{fig2}b).
An upped bound to $\upsilon^2_{\rm max}(\varepsilon)$, valid in the inhomogeneous case ($\alpha_i\neq 0$) and for every $\eps$, 
can be obtained by maximizing each term in Eq.~(\ref{Eq.vH}) separately. This gives
\be  \label{Eq.boundNN}
\upsilon^2_{\rm max}(\varepsilon) \leq
\sum_{i=1}^N \alpha_i^2 + \eps \max\bigg(\sum_{{\rm odd} \, i} \alpha_i , \sum_{{\rm even} \, i} \alpha_i \bigg) + \eps^2 \frac{N}{2},
 \ee
which is shown as dashed red line in Fig.~\ref{fig2}b.
  
It is important to emphasize that different Hamiltonians $H$ detect different subsets of entangled states.   
Linear Hamiltonians ($\eps=0$) are well suited to detect entangled symmetric states. 
Entangled non-symmetric states are better detected by nonlinear Hamiltonians ($\eps\neq 0$). 
Let's consider, for instance, the state of $N$ spins 
$\ket{\chi} = (\ket{\uparrow \downarrow}^{\otimes N/2} + \ket{\uparrow}^{\otimes N/2} \ket{\downarrow}^{\otimes N/2})/\sqrt{2}$.
It is possible to demonstrate that, when applying $e^{-i H_0 \theta}$ with $H_0 = \tfrac{1}{2}\sum_{i=1}^N \sigma_{\vect{n}}^{(i)}$, 
the quantum speed of $\ket{\chi}$ is smaller than the bound $\vmax = N$ for $N >6$, even when optimizing the direction $\vect{n}$. 
Therefore, $\ket{\chi}$ cannot be detected as entangled when probed by only linear Hamiltonians.
Conversely, when probed by a nonlinear nearest-neighbor Hamiltonian $H_1 = \tfrac{1}{4}\sum_{i=1}^N \sigma_{\vect{n}}^{(i)}\sigma_{\vect{n}}^{(i+1)}$, 
this state has a square quantum statistical speed equal to $N^2/4-N+1$ that surpasses the bound $\vmax = N/2$ if $N\geq 6$.
$\ket{\psi}$ can thus be detected as entangled when probed by the Ising Hamiltonian.
The opposite is also true. For instance the Greenberger-Horne-Zeilinger (GHZ) state 
$\ket{\varphi} = (\ket{\uparrow}^{\otimes N} + \ket{\downarrow}^{\otimes N} )/\sqrt{2}$
has a null statistical speed when probed with $H_1$. 
Nevertheless $\ket{\varphi}$ can reach a statistical speed $N^2 > N$ and it can thus be 
detected as entangled when probed by the linear Hamiltonian $H_0= \tfrac{1}{2}\sum_{i=1}^N \sigma_{\vect{n}}^{(i)}$.

Beside the possibility to witness a larger class of entangled states, the measurement of the statistical speed generated
by nonlocal Hamiltonians allows to take in account the residual coupling among neighboring spins that, in contrast, can
limit the experimental implementation of Bell-based DIEWs, in particular when dealing with a large number of ions \cite{BarreiroNATPHYS2013}.
It is also important to notice that the bound (\ref{Eq.statspeed}) is not violated (no false positives are possible) if the state after the unitary transformation is
affected by noise and decoherence (see Fig.~1) that, in full generality, can be modeled 
as a completely-positive trace-preserving map $\Lambda$~\cite{RafalNATCOMM2011, GiovannettiPRL2006}, 
when $\Lambda$ does not depend on $\theta$.


\section{Applications}

Our method to witness entanglement requires to experimentally extract the statistical speed. 
We show below that this can be obtained from the visibility of fringe oscillating as a function of $\theta$,
from moments of the probability distribution or, more generally, by exploiting a basic relation between the statistical speed and the Kullback-Leibler entropy.
We apply our protocols to extract the statistical speed from published data in ions and photons experiments.
In these experiments, the probing Hamiltonian is linear and the above method can be 
extended as a witness of multiparticle entanglement \cite{HyllusPRA2012, TothPRA2012}: the inequality
\begin{equation}  \label{Eq.kentcond}
\upsilon^2 > s k^2 + r^2,
\end{equation}
signals $(k+1)$-partite entanglement (i.e.
among $N$ parties, at least $k$ are entangled), where $s$ is the largest
integer smaller than or equal to $N/k$ and $r = N - sk$. In particular
$\upsilon^2 > (N-1)^2+1$,
obtained from Eq.~(\ref{Eq.kentcond}) with $k=N-1$, is a witness of genuine $N$-partite entanglement. \\


{\bf Statistical speed from dichotomic measurements.}
We consider here the simplest (but experimentally relevant) case
where the measurement results can only take two values, $\mu=\pm 1$.
In this case, Eqs.~(\ref{Eq.Hellinger})~and~(\ref{Eq.Fisher}) simplify to
$\ell^2 = 8 \big[ 1 - \sqrt{ \Po \Pd}- \sqrt{(1- \Po)(1 - \Pd)} \big]$
and 
\be \label{Eq.Fisher.Dicho}
\upsilon^2 = \frac{1}{P_0 (1 - P_0)} \bigg( \frac{\partial P_\theta }{\partial\theta} \Big\vert_{\theta_0}\bigg)^{2},
\ee
respectively, where $\Po \equiv P(+1 \vert \theta_0) $  and $\Pd \equiv P(+1 \vert \theta_0 + \delta \theta) $.
For instance, if 
\begin{equation}  \label{Eq.probability}
P(\pm 1 \vert \theta) = \frac{1 \pm V \cos N \theta}{2},
\end{equation}
where $V$ is the visibility, we can straightforwardly calculate Eq.~(\ref{Eq.Fisher.Dicho}), obtaining
\begin{equation}  \label{Eq.Fishexp}
\upsilon^2 =\frac{V^2 N^2 \sin^{2}(N\theta)}{1-V^{2}\cos^{2}(N\theta)}.
\end{equation}
It is thus possible to detected entanglement when $V > \tfrac{1}{\sqrt{N}}$. 
Notice that, with an increasing number of qubits $N$ the required minimum visibility to detect entanglement decreases. 
Maximally entangled states are detected when 
$V > \sqrt{( 1 - \tfrac{1}{N} )^2 +\tfrac{1}{N^2} }$, that requires a visibility increasing with $N$.\\


{\bf Statistical speed from average moments.}
Not always the probability of different measurement results are available, but only some averaged moments
$\langle \mu \rangle_\theta = \sum_\mu \mu P(\mu \vert \theta)$. 
We can extend the notion of Hellinger distance and statistical speed to the probability distribution $P(\bar{\mu} \vert \theta)$, 
where $\bar{\mu} = \frac{1}{m} \sum_{i=1}^{m} \mu_i$ and $\mu_1, ...\mu_m$ are measurement results.  
We find
$\ell_{\rm mom}^2 = 4 \sum_{\bar{\mu}} ( \sqrt{P(\bar{\mu}\vert \theta_0)} - \sqrt{P(\bar{\mu}\vert \theta_0 + \delta \theta)} )^2$
and 
\begin{equation}  \label{Eq.Fisher.mom}
\upsilon^2_{\rm mom} = \sum_{\bar{\mu}}
\frac{1}{P(\bar{\mu}\vert \theta_0)} \bigg( \frac{d P( \bar{ \mu} \vert \theta)}{d \theta} \Big\vert_{\theta_0} \bigg)^2,
\end{equation}
where the sum extends over all possible values of $\bar{\mu}$.
Using a Cauchy-Schwarz inequality it is possible to demonstrate (see Appendix) that
$\upsilon_{\rm mom} \leq \upsilon \sqrt{m}$.
For $m \gg 1$, the central limit theorem provides
\be \label{Eq.Prob.mom}
P(\bar{\mu} \vert \theta) = \sqrt{\frac{m}{2 \pi (\Delta \mu)_\theta^2}} e^{- \frac{ m (\bar{\mu} - \langle \mu \rangle_\theta)^2}{2 (\Delta \mu)_\theta^2} },
\ee
where $(\Delta \mu)_\theta^2 = \sum_\mu (\mu -\langle \mu \rangle_\theta)^2 P(\mu \vert \theta)$.
To the leading order in $m$, replacing Eq.~(\ref{Eq.Prob.mom}) into Eq.~(\ref{Eq.Fisher.mom}), we obtain 
\be \label{Eq.Prob.mom.CL}
\upsilon^2_{\rm mom} = \frac{m}{(\Delta \mu)_\theta^2} \bigg( \frac{d \langle \mu \rangle_\theta}{d \theta } \Big\vert_{\theta_0} \bigg)^2.
\ee
The entanglement criteria thus becomes $\upsilon^2_{\rm mom}/m > \upsilon_H^2$. 
When probing with linear Hamiltonians, the inequality $\upsilon^2_{\rm mom}/m > s k^2 + r^2$
witness $(k+1)$-partite entanglement from the experimental measurements of average moments. 
These bounds generalize to arbitrary observables the bounds 
to detect entanglement \cite{SorensenNATURE2001} and multipartite entanglement \cite{SorensenPRL2001} 
from the estimation of the mean collective spin \cite{WinelandPRA1994, MaPHYSREP2011}. \\


\begin{figure}[t!]
\includegraphics[width=\columnwidth]{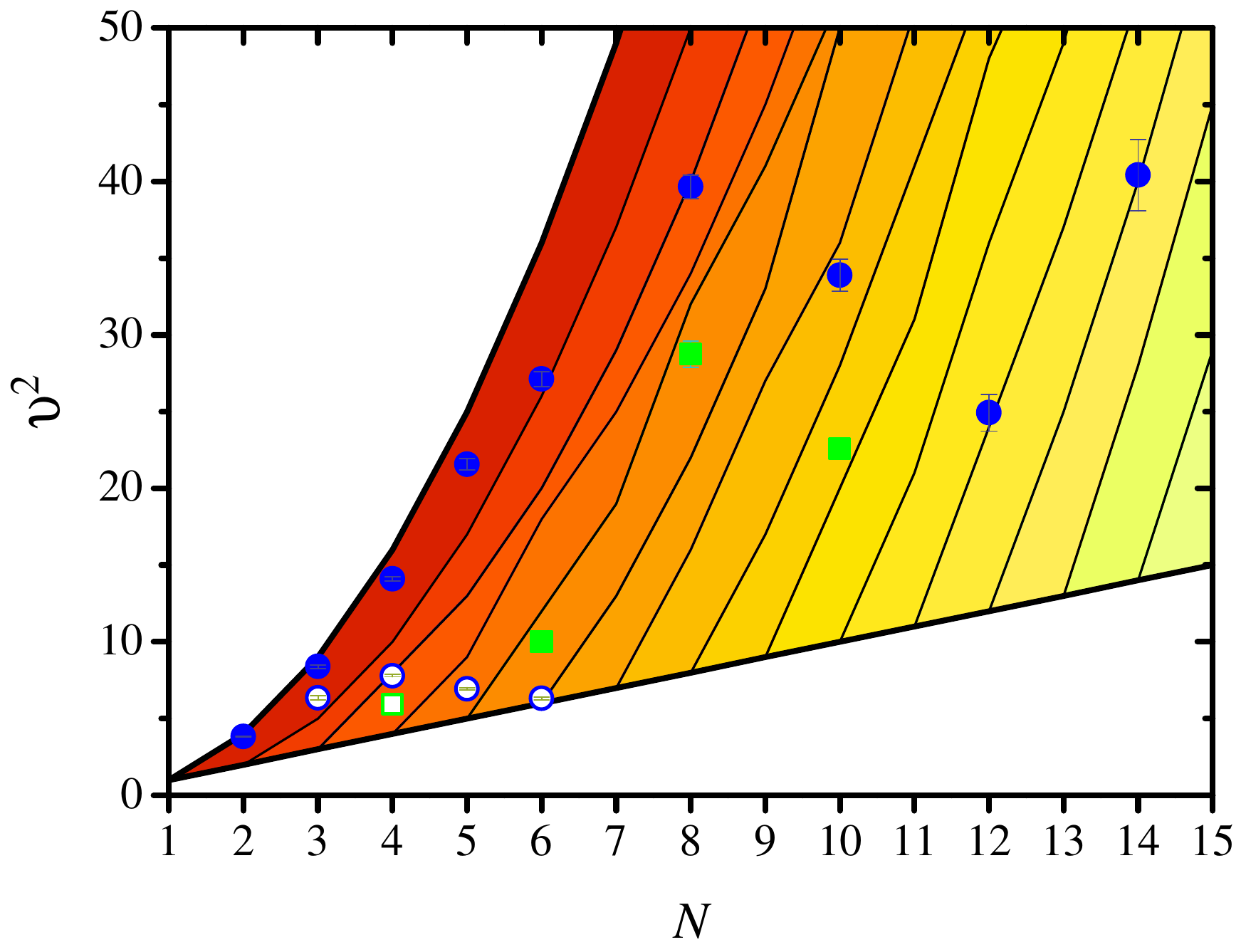}
\caption{\textbf{Witness of multipartite entanglement.} 
Squared statistical speed as a function of the number of qubits obtained analyzing published ions (circles) and photon (squares) experimental data:
Ref.~\cite{MonzPRL2011}, filled circles;
Ref.~\cite{LeibfriedSCIENCE2005} for $N=3$ and Ref.~\cite{LeibfriedNATURE2006} for $N=4,5,6$, open circles;
Ref.~\cite{GaoNATPHYS2012}, filled squares;
Ref.~\cite{WaltherNATURE2004}, open square.
The upper thick line is the upper bound $\upsilon^2=N^2$, the lower thick line is the separability bound $\upsilon^2=N$. 
The different lines are bound for $k$-partite entanglement, Eq.~(\ref{Eq.kentcond}).
In particular, the darker red region stands for genuine $N$-partite entanglement.}
\label{fig3}
\end{figure}

{\bf Witnessing multipartite entanglement in trapped-ions experiments.}
Several recent efforts have been devoted to create a GHZ state
$\tfrac{1}{\sqrt{2}}(\vert 0 \rangle^{\otimes N} + \vert 1 \rangle^{\otimes N})$
with trapped ions
\cite{LeibfriedNATURE2006,LeibfriedSCIENCE2005,MonzPRL2011}.
In Ref.~\cite{MonzPRL2011} the creation of the state has been followed by
a collective rotation $\otimes_{j=1}^N e^{i \tfrac{\pi}{2} \sigma_{\theta}^{(j)} }$, 
with $\sigma_{\theta}^{(j)} = \sigma_{x}^{(j)} \cos{\theta} + \sigma_{y}^{(j)} \sin{\theta}$. 
The output state has been characterized by dichotomic measurements
of the parity, $\Pi =(-1)^{N_0} $, with $N_0$ being the number of qubits measured in one of the two modes.
The reported results are the oscillations of the average parity (and, therefore, of the probability to obtain the $\pm 1$ result) 
as a function of $\theta$, cfr. Eq.~(\ref{Eq.probability}), and we can directly 
analyze the experimental data with our multipartite entanglement witness.
In Fig.~\ref{fig3},  
filled blue circles are obtained from data reported in Ref.~\cite{MonzPRL2011}, for $N=\{2--6,8,10,12, 14\}$,
open blue circles from the data of Ref.~\cite{LeibfriedSCIENCE2005} for $N=3$ and
of Ref.~\cite{LeibfriedNATURE2006} for $N=4,5,6$. 
We first notice that all data satisfy $\upsilon^2>N$: we thus detect entanglement in
\textit{all the states} created in \cite{MonzPRL2011,LeibfriedSCIENCE2005,LeibfriedNATURE2006}. 
The different colored regions correspond to different $k$-partite entanglement detection
[delimited by solid thin lines given by Eq.~(\ref{Eq.kentcond})].
In particular, genuine $N$-partite entanglement is marked by the darker red region that, from the data of Ref.~\cite{MonzPRL2011}, 
is reached up to $N = 6$ ions. 
The maximum value of $\upsilon^2$ is obtained for $N=8$
particles, corresponding to 7-partite entanglement. 
The number of entangled particles in the system slowly decreases for increasing $N$: 
we have $4$-partite entanglement for the states of $N=10$ ions and
$3$-partite entanglement for the state of $N=12$ and $N=14$ ions.
It is interesting to notice that a recent experiment \cite{BarreiroNATPHYS2013} has investigated 
a Bell-based DIEW reporting genuine multipartite entanglement up to $N=6$ ions, which is in agreement with our finding. 
We finally point out that our entanglement witness criteria do not assume any specific state, in particular, not necessarily GHZ-like \cite{Bohnet_2015}. \\


{\bf Witnessing multipartite entanglement in photon experiments.}
Several experiments have demonstrated the creation of multipartite entanglement in photonic systems 
\cite{BouwmeesterPRL1999,ZhaoNATURE2004,GaoNATPHYS2012,YaoNATPHOT2012}. 
In particular, Ref.~\cite{GaoNATPHYS2012} reports on the creation of a GHZ states up to ten photons.
After the creation of the state by parametric down-conversion, a phase shift is applied to each qubit, 
according to the scheme of Fig.~1. 
The state is finally characterized by measuring the operator $\sigma_x^{\otimes N}$, 
whose mean value shows high-frequency oscillations, $\langle \sigma_x^{\otimes N} \rangle = V \cos N \theta$ \cite{GaoNATPHYS2012}.
Noticing that $(\Delta \sigma_x^{\otimes N} )^2 = 1 -  \langle \sigma_x^{\otimes N} \rangle^2$, we can calculate 
the corresponding statistical speed from Eq.~(\ref{Eq.Prob.mom.CL}) to obtain
$\tfrac{\upsilon^2_{\rm mom}}{m} = \frac{V^2 N^2 \sin^{2}(N\theta)}{1-V^{2}\cos^{2}(N\theta)}$.
Also in this case, the witness of multipartite entanglement is solely based on the 
visibility of the interference signal. Results are shown in Fig.~\ref{fig3} (filled squares). 
We witness 4-partite entanglement for the state of $N=8$ photons, giving the highest value of the statistical speed reached with photons.  
As $\upsilon^2_{\rm mom}$ is a lower bound of Eq.~(\ref{Eq.Fisher}), the filled squares in Fig.~\ref{fig3} are lower bound for multipartite entanglement.\\

\begin{figure}[t!]
\includegraphics[width=\columnwidth]{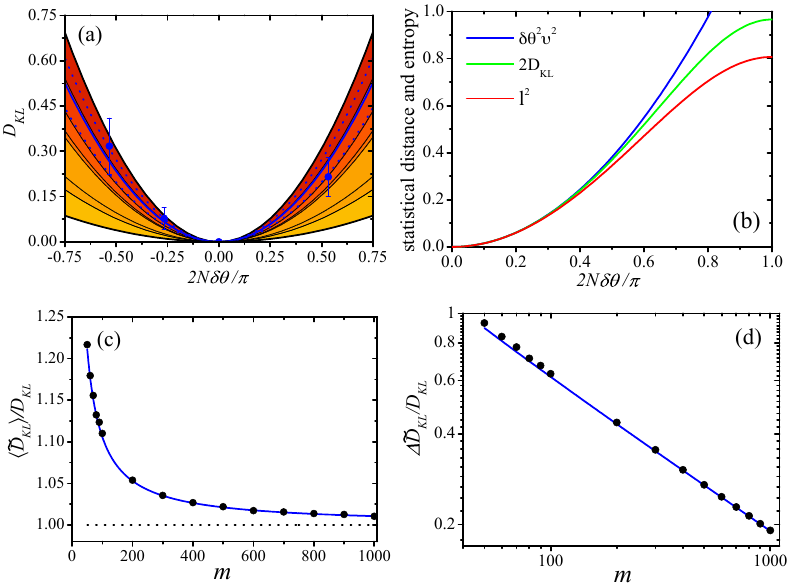}
\caption{
{\bf Kullback-Leibler entropy and statistical speed.}
({\bf a}) KL entropy as a function of $\delta \theta$
obtained from an analysis of the experimental data of Ref.~\cite{MonzPRL2011} for $N=8$ ions.
Blue dots are calculated using Eq.~(\ref{KL}).
The blue line is a parabolic fit, $D_{\rm KL} = \delta \theta^2 \upsilon^2/2$.
Color region corresponds to multipartite entanglement level, with color scale as in Fig.~\ref{fig3}. 
({\bf b}) Squared statistical distance, $\ell^2$ (red line), KL entropy, $2D_{\rm KL}$ (green line), 
and their common low-order approximation, $\delta \theta^2 \upsilon^2$ (blue line), as a function of $\delta \theta$. 
Panels ({\bf c}) and ({\bf d}) reports numerical simulation of
the Kullback-Leibler entropy (dots) as a function of the sample size $m$, for $2N\delta\theta/\pi=0.4$.
Solid lines are analytical predictions, valid for $m \gg 1$ (see text), 
for the statistical bias [in panel (c)] and the statistical fluctuation of $D_{\rm KL}$ [in panel (d)].  
In panels (b)-(d) we used Eq.~(\ref{Eq.probability}) for the probability, 
with $N=8$ and $V=0.787$, consistently with the experimental data of panel (a).}
\label{fig4}
\end{figure}

\textbf{Statistical speed from the Kullback-Leibler entropy.} 
So far we have extracted the statistical speed by fitting the experimental probabilities 
of the different detection events. 
This simple approach can be implemented when the probabilities can be accurately fitted with 
a single parameter function,
as in the ions and photons experiments discussed above.
In general, it might be necessary to extract the statistical speed directly from the bare data without fitting the probability distribution. 
This can be done by experimentally estimating the Kullback-Leibler (KL) entropy \cite{KullbackAMS1951}:
\be \label{KL}
D_{\rm KL} = \sum_{\mu} P(\mu \vert \theta_0) \ln \frac{P(\mu \vert \theta_0)}{P(\mu \vert \theta_0 + \delta \theta)}.
\ee
The KL entropy grows quadratically for small $\delta \theta$
with a coefficient proportional to the squared statistical speed (\ref{Eq.Fisher}) , 
$D_{\rm KL} = \upsilon^2 \delta \theta^2/2$.
In Fig.~\ref{fig4}(a) we illustrate the method using experimental data of Ref.~\cite{MonzPRL2011}.
We focus to the case $N=8$ and calculate $D_{\rm KL}$
around $\theta_0 \approx \pi/(2N)$ according to Eq.~(\ref{KL}).
A quadratic fit provides $\upsilon^2 =44.6 \pm 7.7 $, in agreement with the result ($\upsilon^2 = 39.6 \pm 0.8$)
obtained using Eq.~(\ref{Eq.probability}), see Fig.~\ref{fig3}.
The large error bars are due to the finite sample statistics of the published data
and can be reduced by increasing the sample size and 
concentrating the measurements around a few phase values 
(rather than for the whole $2\pi$ interval). 

To supply to the lack of available experimental data and for illustration purposes, here we implement a numerical Monte Carlo analysis of Eq.~(\ref{KL}) to evaluate the
role of $\delta \theta $ and the sample size, taking, as a testing ground,
the parity measurements with probability (\ref{Eq.probability}). 
In Fig.~\ref{fig4}(b) we plot Eq.~(\ref{KL}) in the case $N=8$ and $V=0.787$, around $\theta_0 \approx \pi/(2N)$. 
The figure shows that the quadratic behavior is
obtained at sufficiently large $\delta \theta $, see also Appendix. 
For comparison, we also show the Hellinger distance as a function of $\delta \theta$
which can also be exploited to extract experimentally the Fisher information \cite{StrobelSCIENCE2014}.
Notice that, for these values of the visibility, due to higher order terms (see Appendix) the quadratic approximation of $D_{\rm KL}$
holds at larger values of $\delta \theta$ than the $\ell^2$ expansion.

A source of noise in the extraction of $\upsilon^2$ is the
limited statistics of the measurement data (other sources of noise, as detection noise and decoherence, result in a reduced visibility). 
Let us indicate as $m$ the
sample size (we consider $m$ measurements performed at phase $\theta _{0}$
and $m$ measurements performed at phase $\theta =\theta _{0}+\delta \theta $). 
The experiment gives access to frequencies rather than probabilities, and
Eq.~(\ref{KL}) extends as $\tilde{D}_{\rm KL} = f_0 \ln \frac{f_0}{f_{\delta \theta}} + (1 - f_0) \ln \frac{1- f_0}{1- f_{\delta \theta}}$, 
where $f_{0}=n_{\mathrm{even,0}}/m$ is
the frequency of even parity results obtained at phase $\theta _{0}$ (and
analogous definition for $f_{\delta \theta }$). Numerically, we can calculate 
$f_{0} $ and $f_{\delta \theta }$ by a Monte Carlo sampling of the probabilities 
$\Po$ and $\Pd$, respectively. In the large-$m$ limit, we can
calculate statistical fluctuations of the Hellinger distance by taking $f_{0}=P_0 + \delta f_{0}$ 
and $f_{\delta \theta }=\Pd+\delta f_{\delta \theta }$,  
and then expanding $\tilde{D}_{\rm KL}$ in Taylor series for small $\delta f_{\delta \theta }$.
We calculate the bias of the squared Hellinger distance, $b\equiv \langle
\tilde{D}_{\rm KL} \rangle -{D}_{\rm KL}$, where brackets indicate statistical
averaging. In the Method section we show that the bias is
positive but decreases as $b\sim 1/2m$ with the sample size $m$. 
Figure~\ref{fig4}(c) shows a comparison between a numerical Monte Carlo analysis and the analytical prediction. 
Similarly, we can evaluate statistical fluctuations
of the KL entropy, $\Delta ^{2}\tilde{D}_{\rm KL} =\langle \tilde{D}_{\rm KL}^{2}\rangle -\langle \tilde{D}_{\rm KL}\rangle ^{2}$. 
We obtain $\Delta^{2} \tilde{D}_{KL} \sim 1/m$, 
showing, also in this case, a scaling inversely proportional to the sample size. 
Details of our analytical calculations are reported in the Method section, a comparison with
analytical calculations is shown in Figure ~\ref{fig4}(d). 
Overall, the simulations show that few hundred measurements are sufficient to extract $\upsilon^2$ with a small 
statistical bias and large signal-to-noise. 


\section{Discussion}

The statistical speed reveals and quantifies entanglement among $N$ parties. 
This requires to probe a quantum state with a generic multi-qubit Hamiltonian.
Our approach shares several important properties of the device-independent entanglement witness based on Bell tests. 
A non-optimal choice of measurement, a noisy implementation of the observable, or
a coupling with a decoherence source affecting the quantum state do not lead to a false detection of entanglement.
It is also not necessary to exactly characterize the Hamiltonian applied to each party: for instance, 
systematic errors in the direction of Pauli matrices are fully tolerated.

The distinguishing property of our protocol is its simplicity: 
both computationally -- it does not need local optimizations depending on the quantum state
-- and experimentally -- it does not require multiple configurations and local operations. 
It also includes generic nonlocal interactions due to experimental tuning  or accidental crosstalk effects. 
The number of operations required to witness entanglement does not increase with the number of parties: the statistical speed is extracted 
from the knowledge of, at least, two probability distributions obtained at nearby values of $\theta$.

We have witnessed $k$-partite and genuine $N$-partite entanglement with trapped ions and photons
by just analyzing published data. 
In particular we have demonstrated genuine $N$-partite entanglement up to $N=6$ ions
in agreement with recent experimental DIEW investigations \cite{BarreiroNATPHYS2013}.
It should be noticed that not all entangled states are characterized by a statistical speed larger than all separable states, even in a noiseless scenario
and optimizing over output measurements. 
Yet, the entangled states violating Eq.~(\ref{Eq.statspeed}) are those (and only those) 
overcoming the maximum interferometric phase sensitivity limit achievable with separable states 
and a phase-encoding transformation $e^{-i H \theta}$.
Similarly, not all entangled states can be
recognized by a Bell inequality: there are (mixed) entangled states that
satisfy all possible Bell's inequalities. 

To conclude, we notice that several experiments have focused on the creation of
GHZ qubit states because, in addition to their foundational interest and possible applications, 
they are recognized by (theoretically) simple witness operators of genuine $N$-partite entanglement.
The method discussed in this manuscript
allows the experimental characterization of a larger class of (hopefully including more robust against decoherence) quantum states. 
Finally, our results show that entanglement can be detected even when the probing Hamiltonian $H$ is nonlinear and therefore generates entanglement. 
This opens the way to study entanglement near quantum
phase transition points by quenching the parameters of the governing many-body Hamiltonian. 

\section{Acknowledgements}

This work was supported by the
National Natural Science Foundation of China (Grant No. 11374197), the PCSIRT (Grant No. IRT13076),
The Hundred Talent Program of Shanxi Province (2012). 


\section{Appendix} 

{\bf Derivation of Eqs.~(\ref{Eq.vH})-(\ref{Eq.vH1}).}
For pure states and unitary transformation $e^{-i \theta H}$, the quantum statistical speed is given by $\vQ(\eps) = 4 (\Delta H)^2$. 
Taking $H = H_0 + \eps H_1$,  $\vQ(\eps)$ is given by Eq.~(\ref{Eq.vH}) with
$\upsilon_0^2 =  4 (\Delta H_0)^2$, 
$\upsilon_1^2 =  4 (\mean{ \{ H_0, H_1\}} - 2 \mean{H_0} \mean{H_1})$ and 
$\upsilon_2^2 =  4 (\Delta H_1)^2$. 
We detail here the calculation of $\upsilon_2^2$ for product pure states,  where $H_1 = \sum_{\substack{i,j=1 }}^N \frac{V_{ij}}{4} \sigma^{(i)}_{\vect{n}} \sigma^{(j)}_{\vect{n}}$.
We have $\upsilon_2^2 = \sum_{i,j,k,l} \tfrac{V_{ij} V_{kl}}{4} 
[\mean{\sigma^{(i)}_{\vect{n}} \sigma^{(j)}_{\vect{n}} \sigma^{(k)}_{\vect{n}} \sigma^{(l)}_{\vect{n}}} - 
\mean{\sigma^{(i)}_{\vect{n}} \sigma^{(j)}_{\vect{n}}} \mean{\sigma^{(k)}_{\vect{n}} \sigma^{(l)}_{\vect{n}}}]$.
Notice that the terms $i\neq j$ and $k\neq l$ do not contribute to $\upsilon_2^2$.
For product states we thus have 
$\mean{\sigma^{(i)}_{\vect{n}} \sigma^{(j)}_{\vect{n}}} \mean{\sigma^{(k)}_{\vect{n}} \sigma^{(l)}_{\vect{n}}} = 
\mean{\sigma^{(i)}_{\vect{n}}} \mean{ \sigma^{(j)}_{\vect{n}}} \mean{\sigma^{(k)}_{\vect{n}}} \mean{ \sigma^{(l)}_{\vect{n}}}$ while 
$\mean{\sigma^{(i)}_{\vect{n}} \sigma^{(j)}_{\vect{n}} \sigma^{(k)}_{\vect{n}} \sigma^{(l)}_{\vect{n}}} =
\mean{\sigma^{(i)}_{\vect{n}}} \mean{ \sigma^{(j)}_{\vect{n}}} \mean{\sigma^{(k)}_{\vect{n}}} \mean{ \sigma^{(l)}_{\vect{n}}}$
only if the indexes $i,j,k,l$ are all different. 
Therefore, only terms where at least two indexes are equal contribute to $\upsilon_2^2$. 
The terms $k=i$, $l=j$ and $k=j$, $l=i$ both contribute with
$\sum_{i,j} \tfrac{V_{ij}^2}{4} (1 - \mean{\sigma^{(i)}_{\vect{n}}}^2 \mean{\sigma^{(j)}_{\vect{n}}}^2 )$.
After straightforward algebra, taking into account all contributing terms, one arrives at Eq.~(\ref{Eq.vH1}).
Repeating the same procedure for $\upsilon_0^2$ and $\upsilon_1^2$, where $H_0 = \sum_{i=1}^N \frac{\alpha_i}{2} \sigma^{(i)}_{\vect{m}}$, 
one derives Eq.~(\ref{Eq.vH}). \\


{\bf Statistical speed for the Ising model.}
We provide here details on the analysis of the Ising model discussed in the main text. 
We consider the homogeneous case $\alpha_i=1$ and $\vect{n}=\vect{m}$.
For $\eps \leq \eps_c$ we find numerically that $\vpr(\eps)$ is maximized 
by taking the same $\mean{\sigma_{\vect{n}}^{(i)}}$ for all $i=1,...,N$, see Fig.~(\ref{fig2}).
The optimization is thus done by replacing $\mean{\sigma_{\vect{n}}^{(i)}}=a$ in Eqs.~(\ref{Eq.vH}) and~(\ref{Eq.vH1}).
This provides the equation 
\be \label{Ising1}
\frac{\vpr(\eps)}{N} = (1-a^2) + 2 \eps (a-a^3) + \frac{\eps^2}{4} (1 + 2 a^2 - 3 a^4)
\ee
that can be maximized over $a$ at fixed value of $\varepsilon$.
The exact analytical expression is long and not reported here. 
For $\eps \ll 1$ we find $a = \eps + O(\eps^3)$, giving $\tfrac{\vmax(\eps)}{N} = 1+  \tfrac{5}{4} \eps^2 + O(\eps^4)$.
For $\eps > \eps_c$ we obtain numerically that $\vpr(\eps)$ is maximized when 
$\mean{\sigma_{\vect{n}}^{(i)}}=1$ and $\mean{\sigma_{\vect{n}}^{(i)}}=0$, giving 
\be \label{Ising2}
\frac{\vmax(\eps)}{N} = \frac{1}{2} + \eps + \frac{\eps^2}{2}.
\ee
Indeed, in the limit $\eps \gg 1$, Eq.~(\ref{Ising2}) goes as $\sim \eps^2/2$ and thus overcomes Eq.~(\ref{Ising1}), which goes as $\sim \eps^2/3$ at best, 
as obtained by maximizing $(1 + 2 a^2 - 3 a^4)$ over $a$.
The value of $\eps$ for which Eq.~(\ref{Ising2}) is equal to the maximum over $a$ of Eq.~(\ref{Ising1}) defines the critical $ \eps_c$. \\ 

\begin{figure}[h!]
\includegraphics[width=\columnwidth]{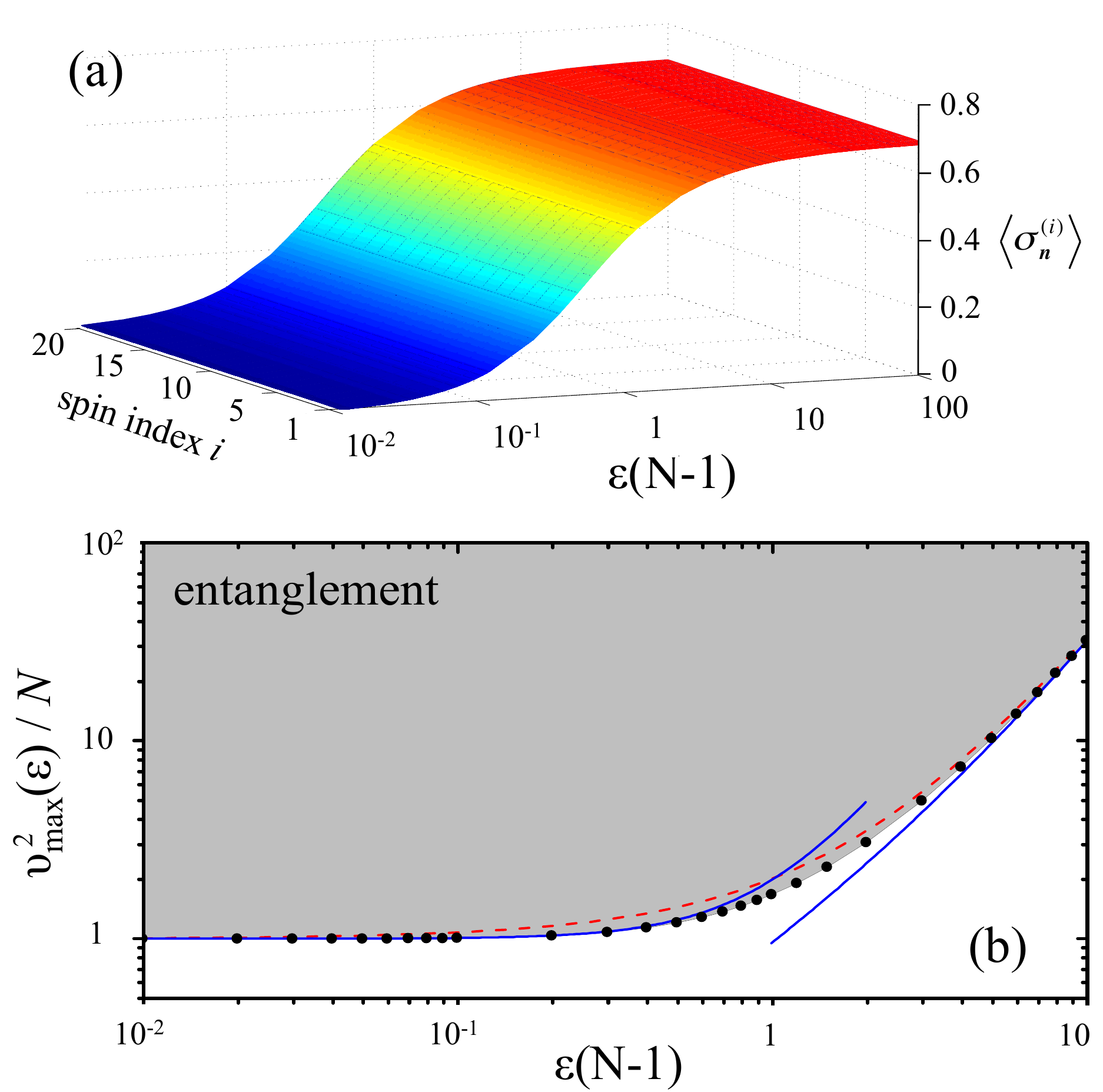}
\caption{\textbf{Witness of entanglement with the LMG Hamiltonian.} 
Same as Fig.~\ref{fig2} but calculated here for the LMG model $V_{ij}=1$. 
({\bf a}) Mean spin $\mean{\sigma_{\vect{n}}^{(i)}}$ of the optimal separable as a function of $\tilde{\eps}$. 
In this case $\mean{\sigma_{\vect{n}}^{(i)}}$ is the same for all spins.
({\bf b}) Dots are the numerical optimization, the solid blue lines are Eq.~(\ref{LMGlow}) for $\tilde{\eps} \equiv \eps (N-1) \ll 1$ and  Eq.~(\ref{LMGhigh}) for $\tilde{\eps} \gg 1$.
The dashed line is the upper bound Eq.~(\ref{Eq.boundLMG}).
}
\label{fig5}
\end{figure}

{\bf Statistical speed for the LMG model.}
We discuss here $\vmax(\eps)$ for the LMG model. 
We consider the Hamiltonian (\ref{H0H1}) with $V_{ij}=1$, $\vect{n}=\vect{m}$ and focus on the homogeneous case $\alpha_i=0$. 
In this case, we numerically find that, for all values of $\varepsilon$, $\vpr(\eps)$ is maximized taking equal $\mean{\sigma_{\vect{n}}^{(i)}}$ for all $i=1,...,N$.
Replacing $\mean{\sigma_{\vect{n}}^{(i)}}=a$ in Eqs.~(\ref{Eq.vH}) and~(\ref{Eq.vH1}) we find
\be \label{LMG1}
\begin{split}
&  \frac{\vpr(\eps)}{N} = (1-a^2) + 2 \tilde{\eps} a (1-a^2) + \\
& \qquad \qquad \qquad + \tilde{\eps}^2 \frac{1 +2 (N-2) a^2 - (2N-1) a^4}{2 (N-1)},
\end{split}
\ee
where $\tilde{\eps} = \eps (N-1)$.
This equation can be maximized over $a$ analytically for each $ \tilde{\eps}$ and $N$ (the explicit expression is long and not reported here).
For $\tilde{\eps} \ll 1$, we have $a = \tilde{\eps}$ giving 
\be \label{LMGlow}
\frac{\vmax(\eps)}{N} = 1+ \tilde{\eps}^2.
\ee
In the opposite $\tilde{\eps} \gg 1$ limit, we find $a = \sqrt{\tfrac{N-2}{2N-1}}$ giving 
\be \label{LMGhigh}
\frac{\vmax(\eps)}{N}  = \tilde{\epsilon}^2 \frac{(N-1)}{2 (2N-3)}
\ee
Maximizing each term of Eq.~(\ref{LMG1}) separately, we can find a upper bound to $\upsilon_H^2$:
\be  \label{Eq.boundLMG}
\frac{\vmax(\eps)}{N}  \leq
    1 +  \frac{4}{3\sqrt{3}}  \tilde{\eps} +\frac{(N-1)}{2(2N-3)} \tilde{\eps}^2.
 \ee
A comparison between numerical results, the limits (\ref{LMGlow}) and (\ref{LMGhigh}), and the bound Eq.~(\ref{Eq.boundLMG}) is shown in Fig.~\ref{fig5}. \\


{\bf Statistical speed for multiple measurements.}
We can extend the notion of Hellinger distance [given in Eq.~(\ref{Eq.Hellinger}) for a single measurement]
to the case of $m$ measurements:
\be
\new{\ell}_{m}^2(\theta_0, \theta) = 8 \Big[ 1-  \sum_{\vect{\mu}}
\sqrt{P(\vect{\mu} \vert \theta_0)  P(\vect{\mu} \vert \theta) } \Big], \nonumber
\ee
where $P(\vect{\mu} \vert \theta)$ is the probability to obtain the sequence $\vect{\mu}\equiv \{ \mu_1, ..., \mu_m \}$
($\mu_i$ is the result of the $i$th measurement)
when the phase shift is equal to $\theta$, and the sum runs over all possible sequences.
For independent measurements we have $P(\mu_1, .., \mu_m \vert \theta) = \prod_{i=1}^m P(\mu_i \vert \theta)$,
and $\ell_{m}^2(\theta_0, \theta)$ becomes
\be
\new{\ell}_{m}^2(\theta_0, \theta) = 8 \Big[ 1 - \Big( \sum_{\mu} \sqrt{P(\mu \vert \theta) P(\mu \vert \theta_0)} \Big)^m \Big]. \nonumber
\ee
A Taylor expansion for $\theta \sim \theta_0$ gives $\ell_m^2 = m \upsilon^2 (\theta-\theta_0)^2$.
In the following we demonstrate that the inequalities 
\be
\new{\ell}_{\rm mom}^2 \leq \new{\ell}_{m}^2, \quad {\rm and} \quad \upsilon^2_{\rm mom} \leq m \upsilon^2 \nonumber
\ee
hold. To show this, let us rewrite the probability of $\bar{\mu}$ as
\be
P(\bar{\mu}\vert \theta) = \sum_{\vect{\mu}}  \delta\big(\xi_m - \bar{\mu}\big) \prod_{i=1}^m P(\mu_i \vert \theta), \nonumber
\ee
where $\xi_m = \tfrac{1}{m} \sum_{i=1}^m \mu_i$ and  
A Cauchy-Schwarz inequality gives
\be
\sqrt{P(\bar{\mu}\vert \theta) P(\bar{\mu}\vert \theta_0)} \geq \sum_{\vect{\mu}}
\delta\big(\xi_m - \bar{\mu}\big)
\prod_{i=1}^m 
\sqrt{P(\mu_i \vert \theta) P( \mu_i \vert \theta_0)} \nonumber
\ee
Summing over $\bar{\mu}$, we conclude that 
\beq
\new{\ell}_{\rm mom}^2 &=& 8 \Big[ 1 - \sum_{\bar{\mu}} \sqrt{P(\bar{\mu}\vert \theta) P(\bar{\mu}\vert \theta_0)} \Big] \nonumber \\
& \leq & 8 \Big[ 1 - \Big( \sum_{\mu} \sqrt{P( \mu \vert \theta) P( \mu \vert \theta_0)} \Big)^m \Big]= \new{\ell}_m^2, \nonumber 
\eeq
From a second-order Taylor expansion of both members around $\theta\sim \theta_0$, we obtain that $\upsilon^2_{\rm mom}/m \leq \upsilon^2$. \\


\textbf{Statistical bias and fluctuations of the Kullback-Leibler entropy.} 
We calculate the statistical bias of the Kullback-Leibler entropy
$b\equiv \langle \tilde{D}_{\rm KL} \rangle -D_{\rm KL} $.
This is done analytically by replacing $f_{\theta }=P_\theta+\delta
f_{\theta }$ into the definition of $\tilde{D}_{\rm KL}$ and performing a
second-order Taylor expansion for $\delta f_{\theta }\ll P_\theta,1-P_\theta$ (we require here $P_\theta\neq 0,1$). We use
binomial statistics, such that $\langle \delta f_{\theta }\rangle =0$, $%
\langle \delta f_{\theta }^{2}\rangle =P_\theta[1-P_\theta]/m$ and
uncorrelated detection, $\langle \delta f_{\theta }\delta f_{0}\rangle =0$.
We obtain
\begin{eqnarray}
b &=&
\frac{P_0}{2m P_{\delta \theta}} +  \frac{1 - P_{0}}{2m (1- P_{\delta \theta})}.
\end{eqnarray} 
Following an analogous method, it is possible to calculate the statistical fluctuations of the KL entropy.
To the leading order in $m$ we have
\be
(\Delta \tilde{D}_{KL})^2  =\frac{\left(  P_{0}-P_{\delta\theta
}\right)  ^{2}}{m(1-P_{\delta\theta})P_{\delta\theta}}
+
\frac{(1-P_{0})P_{0}}{m}
\ln^2 \frac{ (1-P_{0}) P_{\delta \theta} }{ (1-P_{\delta\theta}) P_0 }. 
\ee


\end{document}